\documentclass[aps,prb,twocolumn,superscriptaddress,floatfix,preprintnumbers]{revtex4-1}

\usepackage{amssymb,amsfonts}
\usepackage[intlimits]{amsmath} 
\usepackage[pdftex]{graphicx} 
\usepackage{microtype}
\usepackage{braket}
\usepackage[usenames]{color}
\usepackage{bm}
\usepackage{times}
\usepackage{xspace}
\usepackage[unicode=true,pdfusetitle,
 bookmarks=true,bookmarksnumbered=false,bookmarksopen=false,
 breaklinks=false,pdfborder={0 0 1},backref=false,colorlinks=true,citecolor=blue,
 urlcolor=blue,linkcolor=blue]{hyperref}
\usepackage{physics}
\usepackage{dcolumn}
\usepackage{bm}

\newcommand{\avg}[1]{\left< #1  \right> }

\newcommand{\ff}{\mathcal{N}}

\begin{document}

\title{Incoherent Excitation of Coherent Higgs Oscillations in Superconductors}

\author{Matteo Bellitti}
\affiliation{Department of Physics, Boston University, Boston, MA 02215, USA}
\author{Chris R. Laumann}
\affiliation{Department of Physics, Boston University, Boston, MA 02215, USA}
\author{Boris Z. Spivak}
\affiliation{Department of Physics, University of Washington, Seattle, WA 98195, USA}


\begin{abstract}
We investigate theoretically the excitation of Higgs oscillations of the order parameter in superconductors by incoherent short pulses of light with frequency much larger than the superconducting gap.
The excitation amplitude of the Higgs mode is controlled by a single parameter which is determined by the total number of quasiparticles excited by the pulse.
This fact can be traced back to the universality of the shape of the light-induced quasiparticle cascade at energy below the Debye frequency and above the gap.
\end{abstract}

\maketitle

The dynamics of the superconducting state, described by BCS theory, has been a
subject of research for a long time
\cite{volkov1973collisionless, abrahams1966time, rothwarf1967measurement, chang1978nonequilibrium, papenkort2007coherent, beck2011energy, matsunaga2014light, kemper2015direct, pekker2015amplitude, shimano2020higgs}.
There are two important time scales characterizing the dynamics of the system:
the quasiparticle energy relaxation time $\tau_{in}(\epsilon)$ and $1/\Delta$, where $\epsilon$ is the quasiparticle energy and $\Delta$ is the value of the order parameter.
If the temperature $T$ is not too close to the transition, the quasiparticle energy relaxation rate  is much smaller than the superconducting energy gap:
\begin{equation}
\frac{1}{\tau_{in}(T)}\ll \Delta.
\end{equation}
In this case, for processes with frequency $\omega \ll \Delta$, the density of states is a local function of time and the low frequency dynamics of the superconductor is described by a Boltzmann kinetic equation for the quasiparticle distribution function $n(\epsilon, t)$ and a self-consistent
equation for $\Delta(t)$ (see, \emph{e.g.}, \onlinecite{aronov1981boltzmann}).

At frequency $\omega \sim \Delta$, the Boltzmann approach fails.
Rather, for time $t \ll \tau_{in}$, the dynamics are governed by non-dissipative equations which conserve both the entropy and the total energy.
As a result, the system exhibits coherent oscillations of the order parameter\cite{volkov1973collisionless}, which in the linearized regime decay slowly
\begin{equation}\label{eq:eqOscillations}
\delta \Delta(t)=B\frac{\cos (\omega_{H} t+\phi) }
{\sqrt{\Delta_0 t}}
\end{equation}
Here
$ \omega_H = 2 \Delta_0$  is the Higgs frequency, $ \Delta_0$ is the equilibrium gap, and the parameters $B$ and $\phi$ depend on the initial conditions.

Since the Higgs mode is scalar, it cannot couple linearly to electromagnetic fields directly.
Rather, several excitation mechanisms have been studied:
via combined dynamics of the Higgs mode with charge density wave oscillations \cite{cea2014nature,measson2014amplitude},
linear excitation by coherent THz electromagnetic waves in the presence of DC super-current  \cite{moor2017amplitude,nakamura2019infrared},
and nonlinear coherent excitation  \cite{matsunaga2013higgs,matsunaga2014light} using high intensity THz light.

In this work, we discuss excitation of the Higgs mode by incoherent short light pulses with duration $\tau_{im} \ll 1/\Delta_{0}$ and frequency $\Omega_{0}\gg \Delta_{0}$.
The physical picture of the mechanism is the following:
the pulse creates non-equilibrium quasiparticles with characteristic energy $\epsilon \gg \Delta_0$.
Initially, these quasiparticles are not effective at exciting the Higgs mode because they have high energy and the relaxation rate $1/\tau_{in}$ of their distribution is much faster than $\Delta_{0}$.
As their energy decreases due to various inelastic processes, the relaxation rate decreases as well.
When the typical energy $\epsilon$ is smaller than the Debye energy $\Omega_{D}$ but still much larger than $\Delta_0$, the relaxation is controlled by phonon emission with rate \cite{abrikosov2017fundamentals}
\begin{equation}
    \tau^{-1}_{in}(\epsilon) =\gamma \epsilon^3
\end{equation}
where $\gamma=\alpha/\Omega^{2}_{D}$, and $\alpha$ is a coefficient of order one, just as in a normal metal.

The optimal coupling between the quasiparticle cascade and the Higgs mode is achieved when the rate of change of the quasiparticle distribution function is of order $\omega_H$.
This provides an estimate of the characteristic energy at this stage of the relaxation process,
 \begin{equation}
    \epsilon^\star \equiv ( 2 \Delta_0 \Omega_D^2 /\alpha)^{1/3}
\end{equation}
It is important that $\epsilon^{\star} \gg \Delta_0$:
at such high energy, superconducting correlations are negligible, and we may approximate $n(\epsilon,t)$ with the solution of the Boltzmann equation for a normal metal:
\begin{equation}
  \label{eq:boltzmann}
  \frac{\partial n(\epsilon,t)}{\partial t}
  = \gamma
  \bigg(
  \int_{\epsilon}^{\infty} d\epsilon'   (\epsilon -\epsilon')^2 n(\epsilon',t)
  - \frac{1}{3} \epsilon^3 n(\epsilon,t) \bigg)
\end{equation}
Here, we assume the intensity of the exciting pulse is small and neglect terms nonlinear in $n$ as well as stimulated emission of phonons.

It is furthermore possible to neglect phonon reabsorption during the cascade down to $\epsilon^*$, because the phonon reabsorption rate is much smaller than the inelastic electron relaxation rate
$1/\tau_{ph} \ll 1/\tau_{in}$  in the energy interval $[\epsilon^*, \Omega_D]$.
Indeed, the phonon reabsorption rate is $1/\tau_{ph}\sim \omega c/v_{F}$ at $ql\ll 1$ and $1/\tau_{ph}\sim \omega^{2}\tau$ at $ql \gg 1$. Here $\omega$ and $q$ are the frequency and the wave vector of phonons, $c$ and $v_{F}$ are the speed of sound and the Fermi velocity, $l$ and $\tau$ are the electron mean free path and mean free time.
The desired inequality follows from the fact that $c \ll v_F$.

\begin{figure}
    \centering
    \includegraphics[width=\columnwidth]{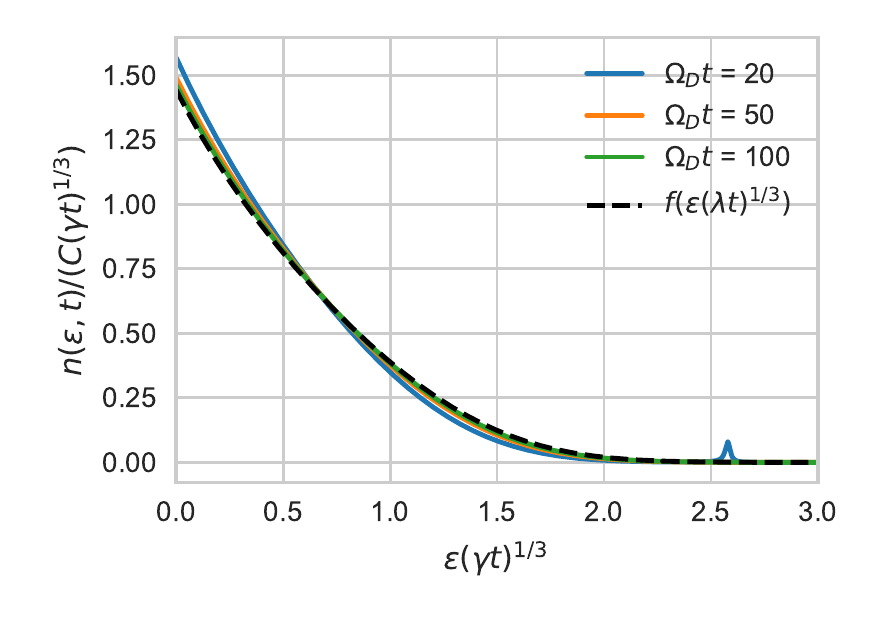}
    \caption{Scaling collapse of the quasiparticle distribution function. The dashed black curve is the solution of Eq.~\eqref{eq:scaling_equation}. The small peak around $\epsilon (\gamma t)^{1/3} = 2.5$ is a signature of the initial conditions: in this particular simulation we started with the quasiparticle population narrowly peaked at $\Omega_D$.}
    \label{fig:scaling_collapse_phonons_loss}
\end{figure}

At $\epsilon \ll \Omega_{0}<\Omega_{D}$ the dynamics described by equation~\eqref{eq:boltzmann} has no scales, and the distribution function approaches a scaling form
\begin{align}
\label{eq:thin_qp_scaling}
    n(\epsilon,t) &= C \Omega_D (\gamma t)^{1/3} \ff \bigg( [\frac{t}{\tau_{in}(\epsilon)}]^{1/3} \bigg)
\end{align}
where $\ff(u)$ is a scaling function satisfying the equation
\begin{align}
  \label{eq:scaling_equation}
   u \ff'(u) + \left( 1 + u^3 \right) \ff(u) = 3 \int_u^{ \infty } du' \, (u - u')^2 \ff(u')
\end{align}
In Eq.~\eqref{eq:thin_qp_scaling}, $C$ is a normalization constant fixing the total number of quasiparticles, which is effectively conserved over the time scale relevant for excitation of the Higgs mode.
We fix $\int du \ff(u) = 1$.
Then $C= N/(\Omega_D \nu(0))$ where $N$ is the number of quasiparticles per unit volume created by the pulse, and $\nu(0)$ is the density of states per unit volume at the Fermi energy. The solution for $\ff$ is shown in Fig.~\ref{fig:scaling_collapse_phonons_loss}.

Let us now turn to the description of the superconducting dynamics induced by the quasiparticle cascade.
In the absence of magnetic field and in the mean field approximation, the non-equilibrium superconducting dynamics at frequencies of order $\Delta$ is described by four equations of motions for Green functions
(normal Keldysh, normal retarded, anomalous Keldysh, anomalous retarded) together with one self-consistent equation for the order parameter $\Delta(t)$.
It has been shown\cite{volkov1973collisionless}, however, that in the uniform, isotropic case and in the absence of inelastic scattering, the four equations for the Green functions  can be reduced to just two for the equal-time Keldysh functions,
\begin{subequations}
\begin{align}
    g(\xi,t) &\equiv \int \frac{d\omega}{2\pi} \, G^K_{11}(t;\xi, \omega) \\
    f(\xi,t) &\equiv \int \frac{d\omega}{2\pi} \, G^K_{12}(t;\xi, \omega)
\end{align}
\end{subequations}
where $\xi = p^2/2m - \epsilon_F$ parameterizes the momentum dependence.
Here, the integrands are
\begin{equation}
\begin{split}
G^K_{\alpha \beta}(t;\mathbf{p},\omega) = - i \int_{-\infty}^\infty d \tau \, e^{i \omega \tau}
\bigg<
    \bigg[
        &\psi_\alpha \bigg(\mathbf{p},t + \frac{\tau}{2} \bigg)
        , \\
        & \psi^\dagger_\beta \bigg(\mathbf{p},t - \frac{\tau}{2} \bigg)
    \bigg]
\bigg>
\end{split}
\end{equation}
with $\alpha,\beta$ Nambu indices.
We have assumed that the elastic relaxation time is the shortest time in the problem so that the Green functions are independent of the direction of $\mathbf{p}$ and can be parametrized by $\xi$ alone, $G^K_{\alpha \beta}(t;\mathbf{p},\omega) = G^K_{\alpha\beta}(t;\xi,\omega)$.

We generalize the equations obtained in Ref.~\cite{volkov1973collisionless} for $g$ and $f$ taking into account inelastic electron-phonon processes in the first order of perturbation theory,
\begin{subequations}
\label{eq:linear_eq20}
\begin{align}
    i \partial_t \delta g + 2 \Delta_0 \text{Re}[\delta\!f] =  I_{11} \\
    \left(i \partial_t - 2 \xi \right) \delta\!f + 2 g_{eq} \delta\Delta + 2 \Delta_0 \delta g =   I_{12} \\
    \delta \Delta = - \frac{ \lambda_\textrm{BCS}}{2}   \int d \xi \, \textrm{Im}[\delta\!f]
\end{align}
\end{subequations}
where $\lambda_\text{BCS}$ is the dimensionless BCS coupling, and
\begin{subequations}
\label{eq:collision_compact_form}
\begin{align}
    I_{11} =  2 \gamma
  \, \textrm{sign}(\xi) \bigg(
   \frac{1}{3} & \epsilon^3 n(\epsilon,t)
 \nonumber \\
  - &\int_{\epsilon}^{\infty} d \epsilon' \,   (\epsilon -\epsilon')^2 n(\epsilon',t)
  \bigg)
    \\
   I_{12} = 2 \gamma \Delta_{0}
    \bigg(
         n(\epsilon,t) &\int_0^{ \epsilon } d \epsilon' \, ( \epsilon - \epsilon' )^2 \left( \frac{ 1}{ \epsilon } - \frac{ 1 }{ \epsilon' }   \right)
         \nonumber \\
        - \int_{  \epsilon  }^{ \infty }  d \epsilon' \, &( \epsilon - \epsilon' )^2 \left( \frac{ 1 }{ \epsilon} - \frac{ 1 }{\epsilon' }   \right)   n(\epsilon',t)
    \bigg)
\end{align}
\end{subequations}
Eqs.~\eqref{eq:linear_eq20} and \eqref{eq:collision_compact_form}  describe the linearized dynamics of small deviations around the equilibrium solution\cite{volkov1973collisionless}
\begin{equation}
\label{eq:fg_equilibrium}
f_{eq}= -i\frac{\Delta_{0}}{\epsilon} (1 - 2 n_F(T))
\quad
g_{eq}= -i \frac{\xi}{\epsilon}(1 - 2 n_F(T))
\end{equation}
where $n_F(T) = (1 + \exp(\epsilon/T))^{-1}$ is the equilibrium Fermi distribution and $\epsilon = \sqrt{\xi^2 + \Delta_0^2}$.
Eq.~\eqref{eq:collision_compact_form} is valid for energies $\epsilon \gg \Delta$, where we need not distinguish the quasiparticle energy $\epsilon$ from $|\xi|$.
We also take into consideration that, for optical excitation, the quasiparticle distribution $n$ is even in $\xi$ and follows the metallic cascade governed by Eq.~\eqref{eq:thin_qp_scaling}.

We sketch details of the derivation of Eqs.~\eqref{eq:linear_eq20} and \eqref{eq:collision_compact_form} in the Appendix.
As we are interested in the case $T\ll \Delta_{0}$, we approximate $1 - 2n_F(T) \simeq 1$.
Furthermore, for optical excitation,  $\textrm{Re}[\delta\!f]$ is odd in $\xi$, while $\textrm{Im}[\delta\!f]$ and $\delta g$ are even in $\xi$ at all times.

\begin{figure}
    \centering
    \includegraphics[width=\linewidth]{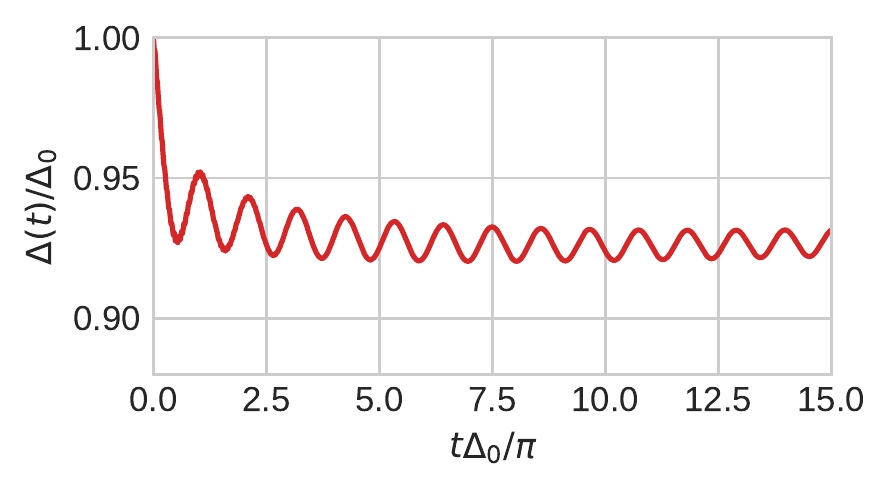}
    \caption{Example solution of Eq.\eqref{eq:linear_eq20}. The units on the horizontal axis are chosen so that a period of one unit corresponds to $\omega = 2\Delta_0$.}
    \label{fig:qualitative_higgs}
\end{figure}

As the excitation of $\Delta(t)$ is controlled by energies $\epsilon\sim \epsilon^{*}\gg \Delta$, we can substitute the solution for the quasiparticle cascade in the normal metal (Eq.~\eqref{eq:thin_qp_scaling}) into the RHS of Eq.~\eqref{eq:collision_compact_form}.
In Fig.\ref{fig:qualitative_higgs} we present an example trace of $\Delta(t)$ with these excitation conditions.

Taking the Fourier transform of Eq.~\eqref{eq:linear_eq20}, solving for $\delta\!f$ in Eqs.~\eqref{eq:linear_eq20}(a,b) and substituting the result into Eq.~\eqref{eq:linear_eq20}c, we obtain
\begin{align}
\label{eq:delta_eom}
    \delta \Delta(\omega) = \frac{1}{i \omega F(\omega)} \bigg( \frac{4 \Delta_0}{\omega^2 - 4 \Delta_0^2} &\left< \frac{\xi I_{11}(\xi,\omega)}{\omega^2 - 4\epsilon^2} \right> \nonumber \\
    & -\left< \frac{I_{12}(\xi,\omega)}{\omega^2 - 4\epsilon^2} \right>    \bigg)
\end{align}
where the angle brackets are shorthand for
\begin{equation}
    \left< \ldots \right> \equiv \frac{ \lambda_\text{BCS}}{ 2} \int_{- \Omega_D}^{ \Omega_D} d \xi \ldots
\end{equation}
and following Ref.~\onlinecite{volkov1973collisionless} we define the auxiliary function
\begin{equation}
    F(\omega) \equiv \left< \frac{1}{\epsilon} \frac{1}{\omega^2 - 4 \epsilon^2} \right>
\end{equation}
For a discussion of the analytic properties of this function see Ref.\cite{volkov1973collisionless}.
The most important feature is the pair of branch points on the real axis at $\omega = \pm 2 \Delta_0$, which is the origin of the $1/\sqrt{t}$ decay of the $\delta \Delta(t)$ oscillation amplitude at long times\cite{lighthill1958introduction} in Eq.~\eqref{eq:eqOscillations}.


%
%
\begin{figure}
    \centering
    \includegraphics[width=0.8\columnwidth]{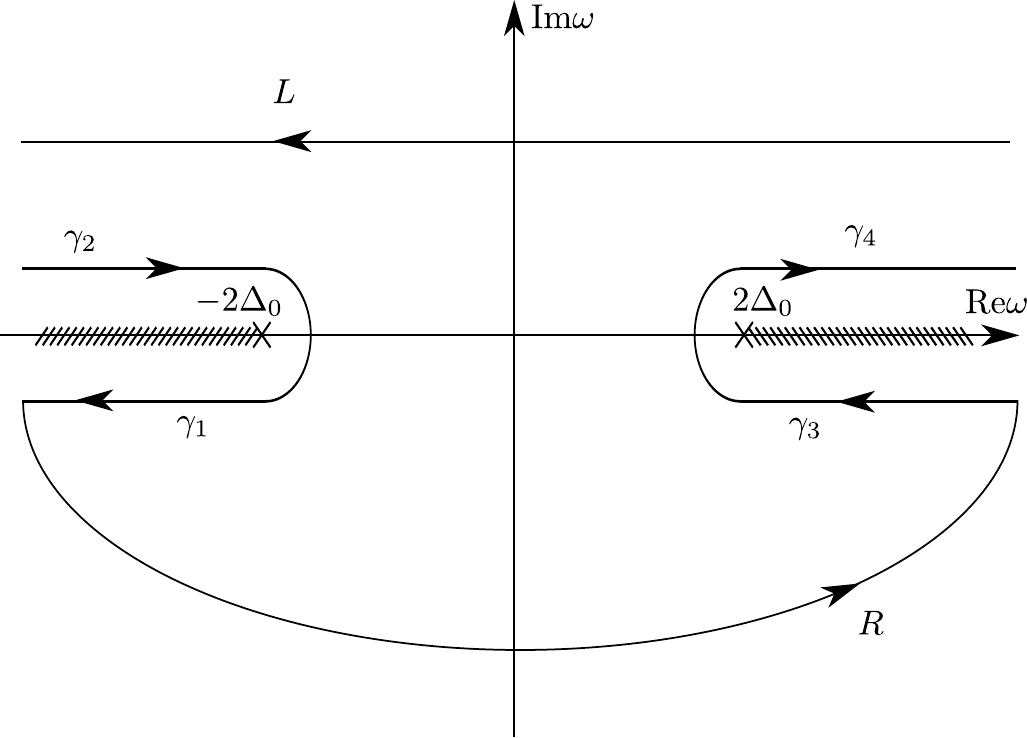}
    \caption{Contour in the complex $\omega$ plane for the solution of Eq.~\eqref{eq:cauchy}. The semicircle $R$ has radius approaching infinity, while the the semicircles surrounding the branch points at $\pm 2 \Delta_0$ have vanishing radius.}
    \label{fig:fourier_contour}
\end{figure}

To obtain the amplitude of oscillations, we take the inverse Fourier transform
\begin{equation}
    \delta \Delta(t) = \int_{-\infty}^\infty \frac{d \omega}{2 \pi} e^{-i\omega t} \delta \Delta(\omega)
\end{equation}
which is equivalent to the integral along the contour $L$ in Fig.~\ref{fig:fourier_contour}.
Using the Cauchy residue theorem we have
\begin{equation}\label{eq:cauchy}
    \delta \Delta(t) = i \Res(\delta \Delta(\omega),0) + \int_{\cup_i \gamma_i} \frac{d \omega}{2 \pi} e^{-i\omega t} \delta \Delta(\omega)
\end{equation}
We focus on the second term in Eq.~\eqref{eq:cauchy} which describes the oscillating part of $\Delta(t)$.
The first term describes the mean value of $\Delta(t)$ at late time, which is determined physically by quasiparticle processes at energies of order the gap $\Delta$ where our perturbative expressions for $I_{11}$ and $I_{12}$ are no longer valid.

Close to the branch points of $F(\omega)$, we have
\begin{equation}
    \left< \frac{4 \Delta_0 \xi I_{11}(\xi,\omega)}{\omega^2 - 4\epsilon^2}\right> \simeq i \pi \frac{\gamma}{4} \frac{C}{\epsilon^\star} \ff(0) a(\omega)
\end{equation}
where $C$ is the normalization constant in Eq.~\eqref{eq:thin_qp_scaling} and
\begin{equation}
    a(\omega) = - \frac{\pi}{\Gamma(-1/3)} + i \frac{\Gamma(1/3)}{6} \text{sign}(\text{Re} \, \omega)
\end{equation}
The dependence on the sign of $\omega$ is due to $I_{11}$ being the Fourier transform of a real function.

It turns out that the term including $I_{12}$ gives a subleading contribution at long times -- the amplitude decays as $1/t$, which is negligible compared to $1/\sqrt{t}$ as $t \to \infty$.
Taking the inverse Fourier transform of Eq.~\eqref{eq:delta_eom}, we arrive at Eq.~\eqref{eq:eqOscillations}
where the values of the parameters $B,\phi$ are
\begin{equation}
 B = \left( \frac{2}{\pi} \right)^{3/2} \frac{\pi |a|  N \ff(0)}{ \nu(0) \Omega_D} \frac{\Omega_D}{\epsilon^\star}
 \qquad
 \phi= \pi/4 + \arg(a)
\end{equation}
where $a = -\pi/\Gamma(-1/3) + i \Gamma(1/3)/6$ so that $|a| \simeq 0.893$. With our choice of normalization, $\ff(0) \simeq 1.44$.

Thus, we arrive at the conclusion that the excitation amplitude of the Higgs mode is controlled by a single parameter which is determined by the total number of quasiparticles excited by a short light pulse with duration $\tau_{im} < 1/\Delta$.
This fact can be traced back to the universality of the shape of the quasiparticle cascade $n(\epsilon,t)$ at energy $\epsilon \sim \epsilon^*$.

\begin{acknowledgements}
CRL acknowledges support from the NSF through grant PHY-1752727.
\end{acknowledgements}

\bibliography{biblio.bib}

\appendix


\section{Derivation of Eq.~\eqref{eq:linear_eq20}}
\label{sec:keldysh_eom}

In this Appendix, we derive the equations of motion for the normal and anomalous Keldysh Green functions using the Keldysh technique~\cite{keldysh1965diagram} in the spatially uniform case.
These were derived in Ref.~\cite{volkov1973collisionless} ignoring the quasiparticle relaxation processes, which we include to leading order in perturbation theory.



We work with the fermionic $G$ and phononic $D$ Green functions in the uniform case,
\begin{align}
\label{eq:gf_fermi_def}
    G^{ij}_{ \alpha \beta}(\bm{p}; t_1, t_2) &\equiv - i \avg{ \mathcal{T}_C \psi_\alpha(\bm{p},t_{1i}) \psi_\beta^\dagger(\bm{p},t_{2j})}
    \\
    D^{ij}(\bm{p}; t_1, t_2) &\equiv - i \avg{ \mathcal{T}_C \phi(\bm{p},t_{1i}) \phi(-\bm{p},t_{2j})}
\end{align}
where $\mathcal{T}_C$ is the contour ordering symbol, $i,j$ are Keldysh indices, and $\alpha,\beta$  are Nambu indices.
Below, after the Keldysh rotation, we denote the retarded, advanced and Keldysh components of the Green functions $G^R, G^A, G^K$ (and similarly for $D, \Sigma$) explicitly.

The equations of motion follow from the matrix Dyson equation.
The Keldysh components yield,
\begin{subequations}
\label{eq:eomgkboth}
\begin{align}
\label{eq:eomgk11}
        i \partial_t G^K_{11} + \Delta G^{K*}_{12} + G^K_{12} \Delta^* = I_{11} \\
\label{eq:eomgk12}
        i \partial_t G^K_{12} - 2\xi  G^K_{12}
        - \Delta G^{K*}_{11}
        + G^K_{11} \Delta
        = I_{12}
\end{align}
\end{subequations}
where $\xi = p^2/(2m) - \epsilon_F$ and
\begin{equation}
\label{eq:twopoint_delta_definition}
    \Delta \equiv \frac{1}{2} \left( \Sigma^{R}_{12} + \Sigma^{A}_{12} \right)
\end{equation}
is the order parameter.
Equation \eqref{eq:eomgkboth} corresponds to Eq.~(16) of Ref\cite{volkov1973collisionless}. The RHS, $I_{11}$ and $I_{12}$, are formally given by
\begin{widetext}
\begin{subequations}
\label{eq:I11_both}
\begin{align}
\label{eq:I11_formal}
  I_{11} &\equiv \Sigma^K_{11} G^A_{11}
    - G^R_{11} \Sigma^K_{11}
    + \Sigma^R_{11} G^K_{11}
    - G^K_{11} \Sigma^A_{11}
    + \Sigma^K_{12} G^{A^\star}_{12}
    + G^R_{12} \Sigma^{K^\star}_{12}
    +
        G^K_{12} \frac{\Sigma^{R\star}_{12}-\Sigma^{A\star}_{12}}{2}
        - \frac{\Sigma^R_{12}-\Sigma^A_{12}}{2} G^{K^\star}_{12}
      \\
\label{eq:I12_formal}
   I_{12} &\equiv \Sigma^K_{11} G^A_{12}
    - G^R_{12} \Sigma^{K\star}_{11}
    - \Sigma^K_{12} G^{A^\star}_{11}
    - G^R_{11} \Sigma^K_{12}
    + \Sigma^R_{11} G^K_{12}
    + G^K_{12} \Sigma^{A^\star}_{11}
    + \frac{\Sigma^R_{12}-\Sigma^A_{12}}{2}
     G^{K^\star}_{11}
     +
    G^K_{11}
    \frac{\Sigma^R_{12}-\Sigma^A_{12}}{2}
\end{align}
\end{subequations}
\end{widetext}
These encode the effect of inelastic processes on the dynamics: $I_{11}$ yields the Boltzmann collision integral in the metal when $\Delta \to 0$ ($I_{12} \to 0$ in this limit).

All operator products in Eq.~\eqref{eq:eomgkboth} and \eqref{eq:I11_both} are space-time contractions.
Also, $\partial_t$ denotes the derivative with respect to average time after changing time coordinates to $t = (t_1 + t_2)/2$ and $\tau = t_1 - t_2$.

We compute the self-energy $\Sigma$ to leading order in the self--consistent Born approximation \cite{abrikosov2012methods}
which gives us
\begin{equation}
\label{eq:self-energy-explicit}
    \Sigma^{ij}_{\alpha \beta}(x,y) = i (-1)^{ \alpha + \beta + i + j }  G^{ij}_{\alpha \beta}(x,y) D^{ij}(x,y)
    \end{equation}
where $x$ and $y$ are spacetime points.

To focus on the equal time dynamics $t = t_1 = t_2$, we take the Fourier transform over $\tau$ of Eqs.~\eqref{eq:eomgkboth}-\eqref{eq:I11_both} and then integrate over the relative frequency.

In order to evaluate $I_{11}$ and $I_{12}$, we take the Green functions in Eq.~\eqref{eq:I11_both} to be in quasiequilibrium form:
the retarded and advanced functions are assumed to always remain in equilibrium BCS form, while the Keldysh ones depend on the average time $t$ only through the quasiparticle distribution function $n(\xi,t)$,
\begin{widetext}
\begin{subequations}
\label{eq:quasiequilibrium_gf}
\begin{align}
    &G^R_{11}(\xi;\omega) = (G^A_{11}(\xi,;\omega))^* = \frac{u_\xi^2}{\omega - \varepsilon + i 0^+} +\frac{v_\xi^2}{\omega + \varepsilon + i 0^+}
    \\[0.5em]
    &G^K_{11}(\xi;\omega,t) = -2 \pi i (1 - 2 n(\xi,t)) \left( u_\xi^2 \delta(\omega - \varepsilon) - v_\xi^2 \delta( \omega + \varepsilon ) \right)
    \\[0.5em]
    &G^R_{12}(\xi;\omega) = (G^A_{12}(\xi;\omega))^* = - \frac{u_\xi v_\xi }{ \omega + \varepsilon + i 0^+ }  + \frac{u_\xi v_\xi}{ \omega - \varepsilon + i 0^+ }
    \\[0.5em]
    &G^K_{12}(\xi;\omega,t) = -2 \pi i u_\xi v_\xi (1 - 2 n(\xi,t)) \left(  \delta( \omega + \varepsilon  ) + \delta( \omega - \varepsilon ) \right)
    \\[0.5em]
\label{eq:phonon_equil_dr} &D^R( k; \omega  ) = (D^A(k; \omega))^* = \abs{m_k}^2 \frac{ \omega_k }{ ( \omega + i0 )^2 - \omega_k^2 }
    \\[0.5em]
    \label{eq:phonon_equil_dk}&D^K( k; \omega, t) = - i \pi \abs{m_k}^2 ( \delta( \omega+ \omega_ k ) + \delta( \omega - \omega_k ))
\end{align}
\end{subequations}
\end{widetext}
where
\begin{align}
 u_\xi^2 = \frac{1}{2} \left( 1 + \frac{\xi}{\varepsilon} \right)
 \qquad
 v_\xi^2 = \frac{1}{2} \left( 1 - \frac{\xi}{\varepsilon} \right)
\end{align}
with $\varepsilon = \sqrt{ \xi^2 + \Delta_0^2}$.
In these expressions, $|m_k|^2$ is the squared electron--phonon matrix element (which is linear in $k$), and $\omega_k = c k$ is the phonon dispersion.
Finally, linearizing the resulting equations, we obtain Eq.~\eqref{eq:linear_eq20}.

\end{document}